\documentclass[preprint,amsmath,amssymb,aps,showkeys,showpacs]{revtex4}
\usepackage[english]{babel}
\usepackage{graphicx}
\usepackage{graphics}
\usepackage{amsmath}
\usepackage{dcolumn}
\usepackage{amssymb}
\usepackage{bm}

\begin{document}
\title{On an electron in an elastic medium with a spiral dislocation}
\author{A. V. D. M. Maia}
\affiliation{Departamento de F\'isica, Universidade Federal da Para\'iba, Caixa Postal 5008, 58051-900, Jo\~ao Pessoa, PB, Brazil.}

\author{K. Bakke}
\email{kbakke@fisica.ufpb.br}
\affiliation{Departamento de F\'isica, Universidade Federal da Para\'iba, Caixa Postal 5008, 58051-900, Jo\~ao Pessoa, PB, Brazil.}

\begin{abstract}

Topological effects of a spiral dislocation on an electron are investigated when it is confined to a hard-wall confining potential. Besides, it is analysed the influence of the topology of the spiral dislocation on the interaction of the electron with a non-uniform radial electric field and a uniform axial magnetic field. It is shown that a discrete spectrum of energy can be obtained in all these cases. Moreover, it is shown that there is one case where an analogue of the Aharonov-Bohm effect for bound states is yielded by the topology of the spiral dislocation.

\end{abstract}

\keywords{spiral dislocation, linear topological defects, Aharonov-Bohm effect, Landau levels}

\maketitle

\section{Introduction}

Dislocations are topological defects that can appear in an elastic medium and can modify its electronic properties. It characterizes the presence of torsion in the elastic medium. Dislocations have a great interest in semiconductors \cite{semi,semi3,semi4} and quantum dots \cite{au}. Examples of them are the screw dislocation and the spiral dislocation \cite{val,put}. One of the ways of describing dislocations is through the Volterra process \cite{kleinert}. It consists in the process of ``cut'' and ``glue'' of the elastic medium. At present days, dislocations have been widely described through the differential geometry. The starting point was the seminal work of Katanaev and Volovich \cite{kat}, where it is shown that the deformation of the elastic medium can be described by the Riemann-Cartan geometry. By following this line of research, quantum effects associated with the topology of a screw dislocation have been reported in several systems, such as an electron in a uniform magnetic field \cite{fur,fur2,fur7,fil}, quantum ring \cite{fur6,fur3,dantas1}, electrons subject to the deformed Kratzer potential \cite{1} and an electron gas in a cylindrical shell \cite{shell}. In the context of the geometric quantum computation \cite{vedral,vedral2}, quantum holonomies for an electron in an elastic medium with a screw dislocation have been discussed in Ref. \cite{bf3}. Topological effects due to a screw dislocation have also been investigated in the harmonic oscillator \cite{fur8,bfb} and in the doubly anharmonic oscillator \cite{b}. On the other hand, based on the Katanaev-Volovich approach \cite{kat}, topological effects on quantum systems that stem from the presence of a spiral dislocation in an elastic medium have just been reported recently. For instance, in studies of geometric quantum phases \cite{bf} and the harmonic oscillator \cite{mb}.

In this work, we go further in the study of quantum systems in an elastic medium with a linear topological defect by focusing on the spiral dislocation \cite{val}. We investigate effects of the topology of a spiral dislocation when an electron is confined to hard-wall confining potential. Moreover, we analyse the effects of the spiral dislocation on the interaction of the electron with a non-uniform radial electric field, and on the Landau quantization \cite{landau}. By searching for bound states solutions to the Schr\"odinger equation in the presence of this topological defect, we also analyse Aharonov-Bohm-type effects \cite{pesk} associated with the topology of the spiral dislocation.

The structure of this paper is: in section II, we introduce the line element that describes the spiral dislocation. Then, we investigate the effects associated with the topology of the spiral dislocation on the confinement of an electron to a hard-wall confining potential; in section III, we analyse the interaction of the electron with a non-uniform radial electric field; in section IV, we discuss the influence of the spiral dislocation on the Landau levels; in section V, we present our conclusions.

\section{Hard-wall confining potential}

As mentioned in the introduction, several works \cite{fur,fur2,fur7,fil,fur6,fur3,dantas1,1,shell} have dealt with the influence of topological defects on a point charge (electrons or holes). Thereby, in this section, we analyse an (spinless) electron subject to a hard-wall confining potential in an elastic medium that possesses a spiral dislocation \cite{val}. This linear topological defect corresponds to the distortion of a circle into a spiral, and it is known as an edge dislocation. The spiral dislocation can be described through the line element \cite{val,bf,mb}:
\begin{eqnarray}
ds^{2}=dr^{2}+2\beta\,dr\,d\varphi+\left(\beta^{2}+r^{2}\right)d\varphi^{2}+dz^{2},
\label{1.1}
\end{eqnarray}
where the constant $\beta$ is the parameter related to the dislocation ($0\,<\,\beta\,<\,1$). The general form of the time-independent Schr\"odinger equation for an electron in the presence of a spiral dislocation is given by \cite{fur,mb} (we shall use the units $\hbar=1$ and $c=1$)
\begin{eqnarray}
\mathcal{E}\psi=-\frac{1}{2m}\left[\frac{1}{\sqrt{g}}\,\partial_{i}\left(\sqrt{g}\,\,g^{ij}\,\,\partial_{j}\right)\right]\psi,
\label{1.2}
\end{eqnarray} 
where $g_{ij}$ is the metric tensor, $g^{ij}$ is the inverse of $g_{ij}$, and thus, $g=\mathrm{det}\left|g_{ij}\right|$.  Moreover, the indices $\left\{i,\,j\right\}$ run over the space coordinates. Therefore, with the spiral dislocation described by Eq. (\ref{1.1}), we have in Eq. (\ref{1.2}):  
\begin{eqnarray}
\mathcal{E}\psi&=&-\frac{1}{2m}\left(1+\frac{\beta^{2}}{r^{2}}\right)\frac{\partial^{2}\psi}{\partial r^{2}}-\frac{1}{2m}\left(\frac{1}{r}-\frac{\beta^{2}}{r^{3}}\right)\frac{\partial\psi}{\partial r}+\frac{\beta}{m\,r^{2}}\frac{\partial^{2}\psi}{\partial r\,\partial\varphi}\nonumber\\
[-2mm]\label{1.3}\\[-2mm]
&-&\frac{1}{2m\,r^{2}}\frac{\partial^{2}\psi}{\partial\varphi^{2}}-\frac{1}{2m}\frac{\beta}{r^{3}}\frac{\partial\psi}{\partial\varphi}-\frac{1}{2m}\frac{\partial^{2}\psi}{\partial z^{2}}.\nonumber
\end{eqnarray}

Observe in Eqs. (\ref{1.1}) and (\ref{1.3}) that this system has the cylindrical symmetry. Hence, the solution to Eq. (\ref{1.3}) can be written in terms of the eigenvalues of the $z$-components of the angular momentum and the linear momentum operators: $\psi\left(r,\,\varphi,\,z\right)=e^{il\varphi+ikz}\,u\left(r\right)$, where $k$ is a constant and $l=0,\pm1,\pm2,\pm3,\pm4\ldots$. By substituting this solution into Eq. (\ref{1.3}), we obtain the following radial equation: 
\begin{eqnarray}
\left(1+\frac{\beta^{2}}{r^{2}}\right)u''+\left(\frac{1}{r}-\frac{\beta^{2}}{r^{3}}-i\frac{2\beta\,l}{r^{2}}\right)u'-\frac{l^{2}}{r^{2}}\,u+i\frac{\beta\,l}{r^{3}}\,u+\left(2m\mathcal{E}-k^{2}\right)u=0.
\label{1.4}
\end{eqnarray}
A possible way of writing the solution to the radial equation (\ref{1.4}) is in the form \cite{mb}:
\begin{eqnarray}
u\left(r\right)=\exp\left(i\,l\,\tan^{-1}\left(\frac{r}{\beta}\right)\right)\times f\left(r\right),
\label{1.5}
\end{eqnarray}
where $f\left(r\right)$ is an unknown function. Then, by substituting the radial wave function (\ref{1.5}) into Eq. (\ref{1.4}), we have 
\begin{eqnarray}
\left(1+\frac{\beta^{2}}{r^{2}}\right)f''+\left(\frac{1}{r}-\frac{\beta^{2}}{r^{3}}\right)f'-\frac{l^{2}}{\left(r^{2}+\beta^{2}\right)}\,f+\tau\,f=0,
\label{1.6}
\end{eqnarray}
where $\tau=2m\mathcal{E}-k^{2}$. From now on, we consider $k=0$ and $\mathcal{E}>0$. Further, let us define $x=2\sqrt{\tau\left(r^{2}+\beta^{2}\right)}$, and thus, we can rewrite Eq. (\ref{1.6}) in the form:
\begin{eqnarray}
f''+\frac{1}{x}\,f'-\frac{l^{2}}{x^{2}}\,f+\frac{f}{4}=0.
\label{1.7}
\end{eqnarray}

Hence, Eq. (\ref{1.7}) corresponds to the Bessel equation \cite{arf,abra}, whose general solution is given by
\begin{eqnarray}
f\left(x\right)=A\,J_{\left|l\right|}\left(x\right)+B\,N_{\left|l\right|}\left(x\right),
\label{1.8}
\end{eqnarray} 
where $J_{\left|l\right|}\left(x\right)$ and $N_{\left|l\right|}\left(x\right)$ are the Bessel functions of first and second kinds \cite{abra,arf}. In order to have a regular solution at the origin, we must take $B=0$ since the Neumann function diverges at the origin. Then, the solution to Eq. (\ref{1.7}) becomes $f\left(x\right)=A\,J_{\left|l\right|}\left(x\right)$.

Now, we are able to analyse the confinement of the electron to a hard-wall confining potential in the elastic medium that contains the spiral dislocation. This confinement is described by imposing that the wave function vanishes at a fixed value of $x$, i.e., it must vanish when $x\rightarrow x_{0}=2\sqrt{\tau\left(r^{2}_{0}+\beta^{2}\right)}$ ($r_{0}$ is fixed): 
\begin{eqnarray}
f\left(x_{0}\right)=0.
\label{1.9}
\end{eqnarray}

With the purpose of observing the influence of the topology of the spiral dislocation, let us assume that $x_{0}\gg1$. Then, we can write \cite{arf,abra}:
\begin{eqnarray}
J_{\left|l\right|}\left(x_{0}\right)\rightarrow\sqrt{\frac{2}{\pi\,x_{0}}}\,\cos\left(x_{0}-\frac{\left|l\right|\,\pi}{2}-\frac{\pi}{4}\right).
\label{1.10}
\end{eqnarray}

Therefore, by substituting (\ref{1.10}) into (\ref{1.9}), we obtain
\begin{eqnarray}
\mathcal{E}_{n,\,l}=\frac{\pi^{2}}{8m\left(r_{0}^{2}+\beta^{2}\right)}\left[n+\frac{\left|l\right|}{2}+\frac{3}{4}\right]^{2},
\label{1.11}
\end{eqnarray}
where $n=0,1,2,\ldots$ is the quantum number related to the radial modes.

Hence, Eq. (\ref{1.11}) corresponds to the allowed energies for an electron subject to a hard-wall confining potential in an elastic medium with a spiral dislocation. The influence of the topology of the spiral dislocation on the energy levels yields an effective radius $\rho_{0}=\sqrt{r_{0}^{2}+\beta^{2}}$. It is worth observing that when the electron is confined to a hard-wall confining potential in the presence of a screw dislocation \cite{fur3}, the angular momentum is modified by the topological effects of the screw dislocation. In this case, the influence of the topological defect on the energy levels is yielded by an effective angular momentum, which gives rise to an Aharonov-Bohm-type effect for bound states \cite{pesk,fur3,fur2}. In contrast to Ref. \cite{fur3}, the influence of the topology of the spiral dislocation does not change the angular momentum quantum number, as we can see in Eq. (\ref{1.11}). In this sense, there is no Aharonov-Bohm-type effect for bound states \cite{pesk,fur3}.

\section{Interaction with a radial electric field}

In this section, let us consider an electric field produced by a uniform distribution of electric charges inside a long conductor cylinder. Besides, there is the presence of the spiral dislocation inside the cylinder. Therefore, the electric field inside the long conductor cylinder is $\vec{E}=\frac{\lambda}{2}\,r\,\hat{r}$, where $\lambda>0$ is a constant associated with the uniform distribution of electric charges. For an electron that interacts with this radial electric field, thus, the potential energy is given by
\begin{eqnarray}
V\left(r\right)=-q\int\vec{E}\cdot d\vec{r}=-q\int E\,dr=\frac{\left|q\right|\,\lambda}{4}\,r^{2},
\label{2.1}
\end{eqnarray}
where $q=-\left|q\right|$ is the electric charge. Then, by following the steps from eq. (\ref{1.2}) to (\ref{1.6}), we have
\begin{eqnarray}
\left(1+\frac{\beta^{2}}{r^{2}}\right)f''+\left(\frac{1}{r}-\frac{\beta^{2}}{r^{3}}\right)f'-\frac{l^{2}}{\left(r^{2}+\beta^{2}\right)}\,f-\frac{m\,\left|q\right|\,\lambda}{2}\,r^{2}\,f+\left(2m\mathcal{E}-k^{2}\right)f=0.
\label{2.2}
\end{eqnarray}

Let us define the parameter $\alpha^{2}=\frac{m\,\left|q\right|\,\lambda}{2}$ and take $k=0$, then, we can make a change of variables given by $\zeta=\alpha\left(r^{2}+\beta^{2}\right)$. Thus, we have in Eq. (\ref{2.2}):
\begin{eqnarray}
\zeta\,f''+f'-\frac{l^{2}}{4\zeta}\,f-\frac{\zeta}{4}\,f+\mu\,f=0,
\label{2.3}
\end{eqnarray}
where $\mu=\frac{1}{4\,\alpha}\left(2m\mathcal{E}+\alpha^{2}\beta^{2}\right)$. Note that when $r\rightarrow\infty$, then, $\zeta\rightarrow\infty$. On the other hand, when $r\rightarrow0$, we have $\zeta=\alpha\,\beta^{2}$. Since $0<\beta<1$, we can consider $\beta^{2}\ll1$. Therefore, when $r\rightarrow0$ we can consider $\zeta\rightarrow0$, without loss of generality. By analysing the behaviour of Eq. (\ref{2.3}) at $\zeta\rightarrow\infty$ and $\zeta\rightarrow0$ \cite{griff,landau,mb}, we can write its solution as 
\begin{eqnarray}
f\left(\zeta\right)=e^{-\frac{\zeta}{2}}\,\zeta^{\frac{\left|l\right|}{2}}\,_{1}F_{1}\left(\frac{\left|l\right|}{2}+\frac{1}{2}-\mu,\,\left|l\right|+1;\,\zeta\right),
\label{2.4}
\end{eqnarray}
where $\,_{1}F_{1}\left(\frac{\left|l\right|}{2}+\frac{1}{2}-\mu,\,\left|l\right|+1;\,\zeta\right)$ is the confluent hypergeometric function \cite{arf,abra}. Observe that the asymptotic behaviour of a confluent hypergeometric function for large values of its argument is given by \cite{abra}
\begin{eqnarray}
\,_{1}F_{1}\left(a,\,b\,;\zeta\right)\approx\frac{\Gamma\left(b\right)}{\Gamma\left(a\right)}\,e^{\zeta}\,\zeta^{a-b}\left[1+\mathcal{O}\left(\left|\zeta\right|^{-1}\right)\right],
\label{2.5}
\end{eqnarray}
therefore, it diverges when $\zeta\rightarrow\infty$. By searching for bound states solutions to the Schr\"odinger equation, we must impose that $a=-n$ ($n=0,1,2,3,\ldots$), i.e., $\frac{\left|l\right|}{2}+\frac{1}{2}-\mu=-n$. With this condition, the confluent hypergeometric function becomes well-behaved when $\zeta\rightarrow\infty$. Then, from the relation $\frac{\left|l\right|}{2}+\frac{1}{2}-\mu=-n$, we obtain
\begin{eqnarray}
\mathcal{E}_{n,\,l}=\sqrt{\frac{2\left|q\right|\,\lambda}{m}}\left(n+\frac{\left|l\right|}{2}+\frac{1}{2}\right)-\frac{\left|q\right|\lambda\,\beta^{2}}{4}.
\label{2.6}
\end{eqnarray}

Hence, Eq. (\ref{2.6}) corresponds to the allowed energies obtained from the interaction of an electron with a radial electric field produced by a uniform distribution of electric charges in the presence of a spiral dislocation. By following the discussion made in the previous section, we have that the topological effects of the spiral dislocation do not provide any contribution to the angular momentum quantum number in the energy levels (\ref{2.6}). Thereby, there is no Aharonov-Bohm-type effect for bound states \cite{pesk,fur2}. The effects of the topology of the spiral dislocation yields a contribution to the energy levels given by the term proportional to $\beta^{2}$, i.e., the second term of the right-hand side Eq. (\ref{2.6}). By taking $\beta=0$ in Eq. (\ref{2.6}), we obtain the spectrum of energy for an electron that interacts with the radial electric field produced by a uniform distribution of electric charges in the absence of defects.

Let us go further by analysing the influence of a hard-wall confining potential on the present system. For a fixed radius $r_{0}$, we have that $\zeta_{0}=\alpha\left(r^{2}_{0}+\beta^{2}\right)$ due to the presence of the spiral dislocation. In this case, the boundary condition (\ref{1.9}) becomes
\begin{eqnarray}
f\left(\zeta_{0}\right)=0.
\label{2.7}
\end{eqnarray}

Let us consider a particular case where the parameter $b=\left|l\right|+1$ has a fixed value and the parameter $\mu$ is large. In this way, for a fixed $\zeta_{0}=\alpha\left(r^{2}_{0}+\beta^{2}\right)$, the confluent hypergeometric function can be written in the form \cite{abra}:
\begin{eqnarray}
_{1}F_{1}\left(a,\,b;\,\zeta_{0}\right)\propto\cos\left(\sqrt{2b\,\zeta_{0}-4a\,\zeta_{0}}-b\frac{\pi}{2}+\frac{\pi}{4}\right).
\label{2.8}
\end{eqnarray}

Therefore, from Eq. (\ref{2.7}), the influence of the hard-wall confining potential on the interaction of the electron with the radial electric yields a discrete spectrum of energy given by
\begin{eqnarray}
\mathcal{E}_{n}=\frac{\pi^{2}}{2m\left(r^{2}_{0}+\beta^{2}\right)}\left[n+\frac{\left|l\right|}{2}+\frac{3}{4}\right]^{2}-\frac{\left|q\right|\lambda\,\beta^{2}}{4}.
\label{2.9}
\end{eqnarray}

Hence, we have two contributions to the energy levels that stems from the topological effects of the spiral dislocation. One of them is the presence of an effective radius $\rho_{0}=\sqrt{r_{0}^{2}+\beta^{2}}$ in analogous way to Eq. (\ref{1.11}). The other contribution is given the term proportional to $\beta^{2}$, i.e., the second term of the right-hand side Eq. (\ref{2.9}). By comparing with Eq. (\ref{1.11}), we also have that the influence of the topology of the spiral dislocation does not change the angular momentum quantum number, therefore, there is no Aharonov-Bohm-type effect for bound states \cite{pesk,fur3}.

\section{Landau quantization}

Let us analyse the interaction of a spinless electron with a uniform magnetic field. It is well-known in the literature that the interaction of an electron with a uniform axial magnetic field yields a discrete spectrum of energy, where each energy level has an infinity degeneracy. This discrete spectrum of energy is called as the Landau levels \cite{landau}. Several studies have been made in the current literature based on the Landau quantization. Some examples are given by Refs. \cite{ll1,ll2,ll3,ll4,ll5,ob3}. In an elastic medium with a disclination and/or a screw dislocation, the effects of these topological defects on the Landau levels have been investigated in Refs. \cite{fur,furtado3,fur6,l1}. In this section, our focus is on the topological effects of the spiral dislocation (\ref{1.1}) on the Landau levels. By following Refs. \cite{fur,furtado3,fur6,l1}, the time-independent Schr\"odinger equation (\ref{1.2}) becomes  
\begin{eqnarray}
\mathcal{E}\psi=-\frac{1}{2m}\frac{1}{\sqrt{g}}\,\left(\partial_{k}-iq\,A_{k}\right)\left[\sqrt{g}\,g^{kj}\left(\partial_{j}-iq\,A_{j}\right)\right]\psi,
\label{3.1}
\end{eqnarray} 
where $A_{k}$ is a covariant component of the electromagnetic 4-vector potential $A_{\mu}=\left(A_{0},\,A_{k}\right)$.

Let us consider an external uniform magnetic field $\vec{B}=B_{0}\,\hat{z}$, where $B_{0}$ is a constant. According to Ref. \cite{l1}, this axial magnetic field can be introduced into the Schr\"odinger equation (\ref{3.1}) through the non-null covariant component of the 4-vector potential, which is given in the form: $A_{\varphi}=\frac{B_{0}\,r^{2}}{2}$. Thereby, Eq. (\ref{3.1}) becomes
\begin{eqnarray}
\mathcal{E}\psi&=&-\frac{1}{2m}\left(1+\frac{\beta^{2}}{r^{2}}\right)\frac{\partial^{2}\psi}{\partial r^{2}}-\frac{1}{2m}\left(\frac{1}{r}-\frac{\beta^{2}}{r^{3}}+iq\,B_{0}\,\beta\right)\frac{\partial\psi}{\partial r}+\frac{\beta}{m\,r^{2}}\frac{\partial^{2}\psi}{\partial r\,\partial\varphi}\nonumber\\
[-2mm]\label{3.2}\\[-2mm]
&-&\frac{1}{2m\,r^{2}}\frac{\partial^{2}\psi}{\partial\varphi^{2}}-\frac{1}{2m}\left(\frac{\beta}{r^{3}}-iq\,B_{0}\right)\frac{\partial\psi}{\partial\varphi}-i\frac{q\,B_{0}\,\beta}{4mr}\,\psi+\frac{q^{2}B_{0}^{2}r^{2}}{8m}\,\psi-\frac{1}{2m}\frac{\partial^{2}\psi}{\partial z^{2}}.\nonumber
\end{eqnarray}

Therefore, the solution to Eq. (\ref{3.2}) can also be given by $\psi\left(r,\,\varphi,\,z\right)=e^{il\varphi+ikz}\,u\left(r\right)$. In this way, we obtain the following radial equation:
\begin{eqnarray}
\left(1+\frac{\beta^{2}}{r^{2}}\right)u''&+&\left(\frac{1}{r}-\frac{\beta^{2}}{r^{3}}-i\frac{2\beta\,l}{r^{2}}+i\,m\,\omega\,\beta\right)u'-\frac{l^{2}}{r^{2}}\,u+i\frac{\beta\,l}{r^{3}}\,u+i\frac{m\omega\beta}{2r}\,u\nonumber\\
[-2mm]\label{3.3}\\[-2mm]
&-&\frac{m^{2}\omega^{2}r^{2}}{4}\,u+\left(2m\mathcal{E}-k^{2}+m\omega\,l\right)u=0,\nonumber
\end{eqnarray}
where $\omega=qB_{0}/m$ is the cyclotron frequency \cite{landau}. Then, let us write the radial wave function $u\left(r\right)$ in the form:
\begin{eqnarray}
u\left(r\right)=\exp\left(i\,\left[l+\frac{m\omega\beta^{2}}{2}\right]\,\tan^{-1}\left(\frac{r}{\beta}\right)\right)\times\exp\left(-i\frac{m\omega\beta\,r}{2}\right)\times h\left(r\right),
\label{3.4}
\end{eqnarray}
where $h\left(r\right)$ is an unknown function. By substituting Eq. (\ref{3.4}) into Eq. (\ref{3.3}), we obtain
\begin{eqnarray}
\left(1+\frac{\beta^{2}}{r^{2}}\right)h''&+&\left(\frac{1}{r}-\frac{\beta^{2}}{r^{3}}\right)h'-\frac{\left(l^{2}+m\omega\,l\,\beta^{2}\right)}{\left(r^{2}+\beta^{2}\right)}\,h-\frac{m^{2}\omega^{2}r^{4}}{4\left(r^{2}+\beta^{2}\right)}\,h\nonumber\\
&+&\left(2m\mathcal{E}-k^{2}+m\omega\,l\right)h=0.
\label{3.5}
\end{eqnarray}

Next, we perform the change of variables $y=\frac{m\omega}{2}\left(r^{2}+\beta^{2}\right)$ and consider $k=0$, and then, Eq. (\ref{3.5}) becomes:
\begin{eqnarray}
y\,h''+h'-\frac{\gamma^{2}}{4y}\,h-\frac{y}{4}\,h+\tau\,h=0,
\label{3.6}
\end{eqnarray}
where we have defined the parameters
\begin{eqnarray}
\gamma&=&l+\frac{1}{2}\,m\,\omega\,\beta^{2};\nonumber\\
[-2mm]\label{3.7}\\[-2mm]
\tau&=&\frac{1}{2m\omega}\left[2m\mathcal{E}+m\omega\,l+\frac{1}{2}m^{2}\omega^{2}\beta^{2}\right].\nonumber
\end{eqnarray}

Observe that the behaviour of Eq. (\ref{3.6}) as $y\rightarrow\infty$ and $y\rightarrow0$ permits us to write the solution to Eq. (\ref{3.6}) as 
\begin{eqnarray}
h\left(y\right)=e^{-\frac{y}{2}}\,y^{\left|\gamma\right|/2}\,_{1}F_{1}\left(\frac{\left|\gamma\right|}{2}+\frac{1}{2}-\tau,\,\left|\gamma\right|+1;\,y\right),
\label{3.8}
\end{eqnarray}
where $\,_{1}F_{1}\left(\frac{\left|\gamma\right|}{2}+\frac{1}{2}-\tau,\,\left|\gamma\right|+1;\,y\right)$ is the confluent hypergeometric function \cite{arf,abra}. By following the steps from Eq. (\ref{2.4}) to Eq. (\ref{2.6}), we obtain
\begin{eqnarray}
\mathcal{E}_{n,\,l,\,k}=\omega\left[n+\frac{\left|\gamma\right|}{2}-\frac{\gamma}{2}+\frac{1}{2}\right].
\label{3.9}
\end{eqnarray}

Hence, Eq. (\ref{3.9}) corresponds to Landau levels in the presence of a spiral dislocation. Note that the topological effects of the spiral dislocation modify  the Landau levels by yielding the parameter $\gamma$ associated with the angular momentum quantum number. In this case, we have a contribution to the angular momentum quantum number in the energy levels. In this sense, there is an analogue of the Aharonov-Bohm effect for bound states. Besides, we have that the degeneracy of the Landau levels is broken due to the topological effects of the spiral dislocation. This change in the degeneracy of the Landau levels is analogous to the effects of topology of the disclination and the screw dislocation on the Landau levels reported in the Refs. \cite{fur,furtado3,fur6}. Finally, observe that it we take $\beta=0$ in Eq. (\ref{3.9}), we recover the Landau levels in the absence of a topological defect \cite{landau}.

\section{Conclusions}

We have investigated the effects of the topology of a spiral dislocation on an electron confined to a hard-wall confining potential. We have obtained a discrete spectrum of energy, where the effects of the topology gives rise to an effective radius $\rho_{0}=\sqrt{r^{2}_{0}+\beta^{2}}$. However, no Aharonov-Bohm-type effect arises from the presence of the topological defect.

In the following, we have analysed the interaction of the electron with a nonuniform electric field in the presence of the spiral dislocation. We have seen that a discrete spectrum of energy can be achieved from this interaction in the elastic medium of the spiral dislocation. The effects of the topology of the defect yield a contribution to the energy levels given by a term proportional to $\beta^{2}$. Besides, we have analysed the influence of a hard-wall confining potential on this system. We have obtained another discrete spectrum of energy for the system, where the topology of the defect gives two contributions to the energy levels. One is the presence of the effective radius $\rho_{0}=\sqrt{r^{2}_{0}+\beta^{2}}$, while the second contribution is a term proportional to $\beta^{2}$. However, in both cases, there is no Aharonov-Bohm-type effect.

Finally, we have analysed the interaction of the electron with an external magnetic field. The interaction of the electron with an axial magnetic field in the presence of the spiral dislocation has gave rise to a discrete spectrum of energy. These energy levels correspond to the Landau levels. We have seen that the influence of the topology of the spiral dislocation modifies the degeneracy of the Landau levels. Moreover, the effects of the topology of the defect yield an effective angular momentum quantum number given by $\gamma=l+\frac{1}{2}\,m\,\omega\,\beta^{2}$ in the Landau levels. Therefore, due to this contribution to the angular momentum quantum number, there is an analogue of the Aharonov-Bohm effect for bound states.

\acknowledgments{The authors would like to thank CNPq for financial support.}

\end{document}